\documentclass{appolb}
\usepackage{graphicx}
\usepackage{amsmath}
\usepackage{scalefnt}
\usepackage{ragged2e}

\begin{document}
\title{The $f_0(1790)$ and $a_0(1950)$ Resonances as Excited $\bar{q}q$ States in the Extended Linear Sigma Model%
\thanks{Presented by D. Parganlija at `Excited QCD 2017', Sintra, Portugal, May 7-13, 2017.}%
}
\author{Denis Parganlija$^a$ \and Francesco Giacosa$^{b,c}$
\address{$^{a}$Institut f\"{u}r Theoretische Physik, 
Technische Universit\"{a}t Wien, Wiedner Hauptstr.\ 8-10, 1040 Vienna, Austria \\
$^{b}$Institute of Physics, Jan Kochanowski University, 
ul.\ Swietokrzyska 15, 25-406 Kielce, Poland \\
$^{c}$Institut f\"{u}r Theoretische Physik, Johann Wolfgang 
Goethe-Universit\"{a}t, Max-von-Laue-Str.\ 1, 60438 Frankfurt am Main, Germany}
}
\maketitle
\begin{abstract}
A decade ago, BES Collaboration reported the discovery of a new scalar isosinglet resonance denoted
as $f_0(1790)$. The finding was subsequently confirmed by LHCb.
Recently, the existence of the corresponding isotriplet
state -- the $a_0(1950)$ resonance -- has been claimed by BABAR.
We investigate whether these resonances can be described as excited $\bar{q}q$ states. To this end, a comprehensive
Lagrangian containing ground-state $\bar{q}q$ mesons as well as their first excitations is constructed in accordance
with symmetries of the strong interaction. Both $f_0(1790)$ and $a_0(1950)$ emerge as compatible with
$\bar{q}q$ excitations; however, tension appears to arise between the simultaneous
interpretation of $f_0(1790)$/$a_0(1950)$ and pseudoscalar mesons
$\eta(1295)$, $\pi(1300)$, $\eta(1440)$ and $K(1460)$ as excited
$\bar{q}q$ states.
\end{abstract}
  
\section{Introduction}
Strong interaction exhibits an abundantly populated spectrum of hadrons. Historical as well as current experimental data
indicate the necessity to introduce 
various quantum numbers for these states -- most notably isospin $I$, total spin $J$, parity $P$ and 
charge conjugation $C$. Mesons are hadrons with integer spin. According to the
Particle Data Group (PDG \cite{PDG}), their number is particularly large in the scalar ($J^{P}=0^{+}$) 
channel where the following resonances are listed in the energy region up to approximately 2 GeV:
\begin{align*}
& f_{0}(500)/\sigma\text{, }K_{0}^{\star}(800)/\kappa\text{, }a_{0}%
(980)\text{, }f_{0}(980)\text{, }f_{0}(1370)\text{, }K_{0}^{\star
}(1430)\text{, }a_{0}(1450)\text{, }\\& f_{0}(1500)\text{, }f_{0}(1710)\text{,
}K_{0}^{\star}(1950)\text{, }a_{0}(1950)\text{, }f_{0}(2020)\text{, }%
f_{0}(2100).
\end{align*}
The abundance is only marginally smaller in the pseudoscalar ($J^{P}=0^{-}$) channel and the same energy region:
\begin{align*}
& \pi\text{, }K\text{, }\eta\text{, }\eta^{\prime}(958)\text{, }\eta
(1295)\text{, }\pi(1300)\text{, }\eta(1405)\text{, }K(1460)\text{, }%
\eta(1475)\text{, }\eta(1760)\text{, }\\& \pi(1800)\text{, }K(1830).
\end{align*}
Masses and decay properties of these states are obviously correlated with their structure; features of
the hadron spectrum can as a matter of principle be explained by the theory of strong interaction
-- Quantum Chromodynamics (QCD). The non-perturbativity of QCD precisely in the energy region where hadrons 
appear \cite{AF1} has brought about the emergence of the famous Quark Model 
and its refined versions (see, e.g., Ref.\ \cite{Rupp}).
In this approach, states are composed
of the constituent quarks -- those emerging from the perturbative quarks of QCD
by means of strong dynamics (see, e.g., Ref.\ \cite{Eichmann:2016yit}). For the states listed above, the expectation
due to their decay patterns is that they are composed of $u$, $d$ and $s$ constituent quarks.
\\\\
The large number of these states implies that not all of them can be explained as having the $\bar{q}q$
(quarkonium) structure -- the spectrum may also contain tetraquark \cite{t} or glueball states \cite{g}. However, the existence of states
with exactly the same quantum number but different masses [$\eta$, $\eta^\prime (958)$, $\eta(1295)$, ...;
$\pi$, $\pi(1300)$, ...; $f_0$ states; ...] leads to an intriguing possibility: that, in addition to ground-state
quarkonia, the meson spectrum may also contain their radial excitations. Here we
explore this further.\\
\\
Studies of excited states (that started already several decades ago \cite{Kataev})
are important for various reasons, for example since the chiral symmetry has been 
suggested to become effectively restored in excited mesons \cite{Wagenbrunn:2006cs} and since
new experimental candidate states have emerged in the last decade.
The observation of the $IJ^{P} = 00^+$ $f_0(1790)$ resonance by the BES and LHCb Collaborations \cite{BES,LHCb}
is of particular
importance. The data suggest the resonance to couple predominantly to pions. The same is also true for lower-lying
resonances [$f_0(500)$, $f_0(1370)$, $f_0(1500)$] where the ground-state quarkonium is expected. Hence it appears
warranted to explore whether $f_0(1790)$ can represent an excited $\bar{q}q$ state.
\\
Recently BABAR \cite{BABAR} has reported the observation of the $IJ^{P} = 10^+$ $a_0(1950)$ resonance; since the ground-state
quarkonium is expected to contribute to the lower-lying $a_0(980)$ and $a_0(1450)$ states, it appears
again warranted to suggest that $a_0(1950)$ represents a $\bar{q}q$ excitation.
We explore these hypotheses by means of the Extended Linear Sigma Model
(eLSM).

\section{Model and Implications}

The Extended Linear Sigma Model is an effective approach to QCD: its degrees of freedom are not quarks
and gluons but rather hadrons. It implements symmetries of QCD as well as their breaking and it contains
degrees of freedom equal to those observed in experiment.
If isospin multiplets are considered single degrees of freedom, there are
16 $\bar{q}q$ ground states and 8 $\bar{q}q$ excited states plus the scalar
glueball in the model. Hence it is expected to entail
important aspects of the strong interaction.
\\
The model has already been used extensively to study $\bar{q}q$ and glueball dynamics in vacuum
\cite{Article2010}. The general form of its Lagrangian is
$
\mathcal{L}=\mathcal{L}_{dil.}+\mathcal{L}_{0}+\mathcal{L}_{E} 
$
where the terms on the right-hand side respectively denote the dilaton (glueball), ground-state $\bar{q}q$
and excited $\bar{q}q$ contributions. $\mathcal{L}_{dil.}$ and $\mathcal{L}_{0}$ are discussed in depth in
Ref.\ \cite{Article2010}. A detailed discussion of the excited-state Lagrangian is presented in
Ref.\ \cite{eMesons}; an abbreviated version is presented in the following.
\\
\\
The excited-state Lagrangian $\mathcal{L}_{E}$ has the following structure \cite{eMesons}:
\begin{align}
\mathcal{L}_{E}  &
=\mathop{\mathrm{Tr}}[(D_{\mu}\Phi_E)^{\dagger}(D_{\mu}\Phi_E)]-(m_{0}%
^{\ast})^{2}\mathop{\mathrm{Tr}}(\Phi
_{E}^{\dagger}\Phi_{E}) +\mathop{\mathrm{Tr}}(\Phi_{E}^{\dagger}\Phi_{E}E_{1}+\Phi_{E}\Phi
_{E}^{\dagger}E_{1}) \nonumber\\
& -\lambda_{2}^{\ast}\mathop{\mathrm{Tr}}(\Phi_{E}^{\dagger}%
\Phi_{E}\Phi^{\dagger}\Phi+\Phi_{E}\Phi_{E}^{\dagger}\Phi\Phi^{\dagger
}) 
-\xi_{2}\mathop{\mathrm{Tr}}(\Phi
_{E}^{\dagger}\Phi\Phi_{E}^{\dagger}\Phi+\Phi^{\dagger}\Phi_{E}\Phi^{\dagger
}\Phi_{E}) \nonumber\\
&  +h_{2}^{\ast}\mathop{\mathrm{Tr}}(\Phi_{E}^{\dagger}L_{\mu}L^{\mu}\Phi
+\Phi^{\dagger}L_{\mu}L^{\mu}\Phi_{E}+R_{\mu}\Phi_{E}^{\dagger}\Phi R^{\mu
}+R_{\mu}\Phi^{\dagger}\Phi_{E}R^{\mu})\nonumber\\
&  +2h_{3}^{\ast}\mathop{\mathrm{Tr}}(L_{\mu}\Phi_{E}R^{\mu}\Phi^{\dagger
}+L_{\mu}\Phi R^{\mu}\Phi_{E}^{\dagger}) -\kappa_{2}%
[\mathop{\mathrm{Tr}}(\Phi_{E}^{\dagger}\Phi+\Phi^{\dagger}\Phi_{E}%
)]^{2}  \text{.}
\label{LagrangianE}%
\end{align}
It is constructed under the conditions that (\textit{i}) the chiral and dilatation symmetries
(and the breaking mechanism as appropriate)
are considered; (\textit{ii}) any terms that lead to mixing of Lagrangian states or
terms suppressed in the limit of large number of colours ($N_c$) are neglected\footnote{The
only exception to this condition is the $\kappa_2$ term that is
necessary to induce the mass splitting of $f_0(1790)$ and $a_0(1950)$, see also the discussion below.}
and (\textit{iii}) only terms that turn out to lead to kinematically allowed decays are included.
\\
In Eq.\ (\ref{LagrangianE}), $\Phi_{E}$ is the multiplet containing excited $\bar{q}q$ states. For three
flavours it reads $\Phi_{E} =\sum_{i=0}^{8}(S_{i}^E+iP_{i}^E)T_{i}$ where
$T_{i}\,(i=0,\ldots,8)$ denote the generators of $U(3)$, while $S_{i}^E$
and $P_{i}^E$ are respectively the scalar and pseudoscalar fields. Then we have
\begin{equation}
\Phi_{E}
=\frac{1}{\sqrt{2}}\left(
\begin{array}
[c]{ccc}%
\frac{(\sigma_{N}^{E}+a_{0}^{0E})+i(\eta_{N}^{E}+\pi^{0E})}{\sqrt{2}} &
a_{0}^{+E}+i\pi^{+E} & K_{0}^{\star+E}+iK^{+E}\\
a_{0}^{-E}+i\pi^{-E} & \frac{(\sigma_{N}^{E}-a_{0}^{0E})+i(\eta_{N}^{E}%
-\pi^{0E})}{\sqrt{2}} & K_{0}^{\star0E}+iK^{0E}\\
K_{0}^{\star-E}+iK^{-E} & {\bar{K}_{0}^{\star0E}}+i{\bar{K}^{0E}} & \sigma
_{S}^{E}+i\eta_{S}^{E}%
\end{array}
\right)  \text{.} \label{PhiE}%
\end{equation}
$\Phi$ is the multiplet containing ground-state scalars and pseudoscalars. $L_\mu$
and $R_\mu$ are the multiplets containing ground-state vectors and axial-vectors;
the structure of these matrices is analogous to that of $\Phi_E$. Additionally,
$D_{\mu}\Phi_E = \partial^{\mu}\Phi^{E}-ig_{1}^{E}(L^{\mu}\Phi^{E}%
-\Phi^{E}R^{\mu})$ is the derivative transforming covariantly under the chiral $U(3) \times U(3)$ group.
Non-vanishing quark masses induce explicit breaking of the chiral symmetry, modelled here via the Lagrangian 
term containing $E_{1}=$ diag$\{0,0,\epsilon_{S}^{E}\}$. Note that the spontaneous chiral-symmetry breaking
is implemented in the ground-state sector via shifting the non-strange and strange $IJ^P = 00^+$ fields 
by their respective vacuum expectation values.
\\
\\
In accordance with our hypotheses, the excited non-strange $IJ^P = 00^+$ state $\sigma_N^E$ is assigned
to $f_0(1790)$; its isotriplet partner $a_0^E$ is assigned to $a_0(1950)$. The non-strange and strange
$IJ^P = 00^-$ states $\eta_N^E$ and $\eta_S^E$ are assigned to the $\eta(1295)$ and $\eta(1440)$ 
resonances\footnote{As
discussed in Ref.\cite{eMesons}, there
is uncertainty whether the energy region $\sim 1.4$ GeV contains one or two pseudoscalar states: PDG
listings \cite{PDG} contain $\eta(1405)$ and $\eta(1475)$ while only $\eta(1440)$ appears
in BES data \cite{BES1998}. Here, $\eta(1440)$ is present
but our results would remain virtually unchanged if, alternatively, $\eta(1475)$ data were used.}.
With this, parameters (or parameter combinations \cite{eMesons}) entering all mass terms
can be calculated.
Four masses are predicted -- and all values can be found in Table~\ref{Decayst3}.
\\
\\
Current experimental situation allows only the determination of parameters relevant for decays of excited
into ground states.
Nonetheless,
just two parameters -- $h_{2}^{\ast}$ and $h_{3}^{\ast}$, fixed from $\Gamma_{f_{0}(1790)\rightarrow\pi\pi} = (270 \pm 45)$ MeV
and $\Gamma_{f_{0}(1790)\rightarrow KK} = (70 \pm 40)$ MeV \cite{BES} --
lead to a prediction of more than 35 decays for almost all other model states.
See Table~\ref{Decayst3} for all numbers.
\\
\\
The results are summarised as follows (for more details, see Ref.\ \cite{eMesons}):

\begin{itemize}
 \item The excited states are generally rather narrow. An exception is the result for the $f_0(1790)$ and $\eta(1440)$.
 Nonetheless, $\Gamma_{f_0(1790)}$ is compatible with the LHCb data
 \cite{LHCb}. The large interval for the $\eta(1440)$ width is a consequence of 
 parameter uncertainties induced by the large errors for $\Gamma_{f_{0}(1790)\rightarrow\pi\pi}$ and
 $\Gamma_{f_{0}(1790)\rightarrow KK}$, see above. These uncertainties also lead to extremely large errors
 [$\cal O$(1 GeV)] of the decay widths of the excited pion and kaon. 
 Hence these states are omitted from Table~\ref{Decayst3}.\\
 We note, however, that if the excited pseudoscalars in the model are implemented to reproduce
 exactly the data on the putative experimental candidates [$\eta(1295)$, $\pi(1300)$, $\eta(1440)$,
 $K(1460)$] then all excited scalars become unmeasurably broad [widths $\cal O$(1 GeV)]. Although
 this result is based on at times ambiguous experimental input and hence care is needed in its
 interpretation (see Ref.\ \cite{eMesons}), it appears to 
 reveal tension between the simultaneous
interpretation of $f_0(1790)$/$a_0(1950)$ and $\eta(1295)$, $\pi(1300)$, $\eta(1440)$ and $K(1460)$ as excited
$\bar{q}q$ states.
 
 \item Our results predict $\Gamma_{a_0^E} = (280 \pm 90)$ MeV;
 this overlaps fully with $\Gamma_{a_0(1950)} = (271 \pm 40)$ MeV measured by BABAR \cite{BABAR}. Hence
 $a_0(1950)$, if confirmed, represents a very good candidate for an excited $\bar{q}q$ state.
 
 \item For $\eta(1295)$, the three decay widths accessible
 to our model (for $\eta_N^E \rightarrow \eta \pi \pi + \eta^{\prime} \pi \pi + \pi KK$) amount
 to $(7 \pm3)$ MeV and hence contribute very little to the overall decay width $\Gamma_{\eta(1295)}^{\text{total}} = (55 \pm 5)$ MeV.
 
 \item Our $\bar{s}s$ scalar isosinglet state $\sigma_S^E$ has the same quantum numbers
 as
 the (unestablished \cite{PDG}) resonances $f_0(2020)$ and $f_0(2100)$ but there is no mass/width
 overlap. Hence they do not appear to represent unmixed excited quarkonia.
 The opposite is true for the (again unestablished) $K_{0}^{\star} (1950)$ resonance: 
 since $m_{K_0^{\star}(1950)} = (1945 \pm 22)$ MeV and $\Gamma_{K_0^{\star}(1950)} = (201 \pm 90)$ MeV \cite{PDG}, 
 it has a
 significant overlap with our excited scalar kaon $K_{0}^{\star E}$.
 
 \end{itemize}

\section{Conclusion}
 
Results from the Extended Linear Sigma Model (eLSM) indicate that the $f_0(1790)$ and -- if confirmed --
also the $a_0(1950)$ and $K_{0}^{\star} (1950)$ resonances are largely unmixed excited $\bar{q}q$ states. The same is
quite likely for $\eta(1295)$ and $\eta(1440)$ although overall, based on the current data, 
there appears to be
tension between the simultaneous
interpretation of $f_0(1790)$/$a_0(1950)$ and $\eta(1295)$, $\pi(1300)$, $\eta(1440)$ and $K(1460)$ as excited
$\bar{q}q$ states. Uncertainties in these conclusions come from (\textit{i}) possible glueball admixture and
(\textit{ii}) scarcity of experimental data that can hopefully be amended by PANDA \cite{PANDA} 
and NICA \cite{NICA}.

\begin{table}[h]
\begin{tabular}{|c|c|c|c|c|}\hline
Excited state
& 
$IJ^P$
& 
Mass (MeV)
&
Decay
&
Width (MeV)
  
\tabularnewline\hline

$f_0(1790)$ & $00^+$ & $1790 \pm 35$*
&
\begin{tabular}{c}
$\sigma_N^E \rightarrow \pi \pi$   \\\hline  
 $\sigma_N^E \rightarrow KK$  \\\hline 
 $\sigma_N^E \rightarrow a_1(1260) \pi$ \\\hline
 $\sigma_N^E \rightarrow \eta \eta^{\prime}$ \\\hline
 $\sigma_N^E \rightarrow \eta \eta$ \\\hline
 $\sigma_N^E \rightarrow f_1(1285) \eta$ \\\hline
 $\sigma_N^E \rightarrow K_1 K$ \\\hline
 $\sigma_N^E \rightarrow \sigma_N \pi \pi$\\\hline
 Total
   \end{tabular}
&
\begin{tabular}{c}
 $\; \, 270 \pm 45$*   \\\hline
 $\; \, \; \; 70 \pm 40$*   \\\hline
 $\; \, 47 \pm \; \, 8$   \\\hline
 $\; \, 10 \pm \; \, 2$   \\\hline
 $\; \;  \, 7 \pm \, \, 1$   \\\hline

 $\; \; \; \, 1 \pm \; \, 0$   \\\hline
 $ \; \; 0 $   \\\hline
 $ \; \;  0 $ \\\hline
 $405 \pm 96$
   \end{tabular}

\tabularnewline\hline
$a_{0}(1950)$ & $10^+$ & $1931 \pm 26$*
&
\begin{tabular}{c}
 $a_{0}^{E} \rightarrow \eta \pi$   \\\hline  
 $a_{0}^{E} \rightarrow KK$  \\\hline 
 $a_{0}^{E} \rightarrow \eta^{\prime} \pi$ \\\hline
 $a_{0}^{E} \rightarrow f_1(1285) \pi$ \\\hline
 $a_{0}^{E} \rightarrow K_1 K$  \\\hline
 $a_{0}^{E} \rightarrow a_1(1260) \eta$ \\\hline
 $a_{0}^{E} \rightarrow a_0(1450) \pi \pi$ \\\hline
 Total
   \end{tabular}
&
\begin{tabular}{c}
 $\; \; 94 \pm 16$   \\\hline
 $\; \; 94 \pm 54$   \\\hline
 $\; \, 48 \pm \; \, 8$   \\\hline
 $\; \, 28 \pm \; \, 5$   \\\hline
 $\; \; \; \, 9 \pm \; \, 5$ \\\hline
 $\; \; \; \, 6 \pm \; \, 1$   \\\hline
 $\; \; \; \, 1 \pm \; \, 1$ \\\hline
 $280 \pm 90$
   \end{tabular}

 \tabularnewline\hline

$\eta(1295)$ & $00^-$ & $ 1294 \pm\, \,4 $*
&
$\eta_N^E \rightarrow \eta \pi \pi + \eta^{\prime} \pi \pi + \pi KK$
&
 $\; \; \; \, 7 \pm \; \, 3$   

 \tabularnewline\hline

$\eta(1440)$ & $00^-$ & $1432 \pm 10$*
&
\begin{tabular}{c}
 $\eta_{S}^{E} \rightarrow K^{\star} K$   \\\hline  
 $\eta_{S}^{E} \rightarrow KK \pi$  \\\hline 
 $\eta_{S}^{E} \rightarrow \eta \pi \pi$ and $\eta^{\prime} \pi \pi$  \\\hline
 Total
   \end{tabular}
&
\begin{tabular}{c}
 $\; \, 128 ^{+204}_{-128}$   \\\hline
 $\; \; \; \, 28 ^{+41}_{-28} \; \,$   \\\hline
 suppressed   \\\hline
 $\; \, 156 ^{+245}_{-156}$
   \end{tabular}
   
\tabularnewline\hline

\begin{tabular}{c}
 $\sigma_S^E$   \\
 (no assignment  \\
  since no $\;\;\;\;\;\;\;\;\;$ \\
  experimental $\;$ \\ candidate $\;\;\;\;\;\;\,$ \\ with congruent \\ mass/width) $\;$ 
 \end{tabular} & $00^+$ & $2038 \pm 24 \;$
&
\begin{tabular}{c}
 $\sigma_S^E \rightarrow KK$  \\\hline 
 $\sigma_S^E \rightarrow \eta \eta^{\prime}$ \\\hline
 $\sigma_S^E \rightarrow \eta \eta$ \\\hline
 $\sigma_S^E \rightarrow K_1 K$ \\\hline
 $\sigma_S^E \rightarrow \eta^{\prime} \eta^{\prime}$ \\\hline
 $\sigma_S^E \rightarrow \pi \pi$, $\rho \rho$ and $\omega \omega$    \\\hline 
 $\sigma_S^E \rightarrow a_1(1260) \pi$ and $f_1(1285) \eta$ \\\hline
 $\sigma_S^E \rightarrow \pi^E \pi$ and $\eta_N^E \eta$ \\\hline
 $\sigma_S^E \rightarrow \sigma_S \pi \pi$\\\hline
 Total
   \end{tabular}
&
\begin{tabular}{c}
$\; \; \; \, 24 ^{+46}_{-24} \; \,$   \\\hline
 $\, \, 16 \pm \, 3$   \\\hline
 $\; \, \, 7 \pm 1$   \\\hline
 $\; \, \, 4^{+8}_{-4}$ \\\hline
 $\; \, \, 1 \pm 0$   \\\hline
 suppressed   \\\hline
 suppressed   \\\hline
 suppressed \\\hline
 suppressed \\\hline
 $\; \; \; \, 52 ^{+58}_{-32} \; \,$
   \end{tabular}

\tabularnewline\hline

\begin{tabular}{c}
$K_{0}^{\star E}$  \\
 $[$tentatively \\ assigned $\;\;\;$ \\ to the $\;\;\;\;\;\;$ \\ $\;\;$unconfirmed \\ 
 $K_{0}^{\star} (1950) \;\;\,$ \\ resonance$]\;$ 
   \end{tabular}
 & $\frac{1}{2}0^+$ & $2023 \pm 27\;$
&
\begin{tabular}{c}
 $K_{0}^{\star E} \rightarrow \eta^{\prime} K$   \\\hline  
 $K_{0}^{\star E}\rightarrow K \pi$  \\\hline 
 $K_{0}^{\star E} \rightarrow K_1 \pi$ \\\hline
 $K_{0}^{\star E} \rightarrow a_1(1260) K$ \\\hline
 $K_{0}^{\star E} \rightarrow \eta K$ \\\hline
 $K_{0}^{\star E} \rightarrow f_1(1285) K$  \\\hline
 $K_{0}^{\star E} \rightarrow K_1 \eta$  \\\hline
 $K_{0}^{\star E} \rightarrow K_0^{\star}(1430) \pi \pi$ \\\hline
 Total
   \end{tabular}
&
\begin{tabular}{c}
 $\; \; 72 \pm  12$   \\\hline
 $\; \; 66 \pm 46$   \\\hline
 $\; \; 10 \pm \; \, 7$   \\\hline
 $\; \; \; 6 \pm \, 4$   \\\hline
 $\; \; \; 6^{+9}_{-6}$   \\\hline
 $\; \; \; \, 2 \pm \; 1$ \\\hline
 $ \; \; 0 $ \\\hline
 $ \; \; 0 $ \\\hline
 $\,  162 ^{+79}_{-76}$
   \end{tabular}

\tabularnewline\hline

\end{tabular}
\caption{Masses and decays of the excited $\bar{q}q$ states. Widths marked as ``suppressed''
depend only on large-$N_c$ suppressed parameters that have been set to
zero. Masses/widths marked with (*) are used as input; the others are predictions.}
\label{Decayst3}%
\end{table}

\textbf{Acknowledgments.}
We are grateful to D.~Bugg, C.~Fischer and A.~Rebhan for extensive discussions. Collaboration with Stephan 
H\"{u}bsch within a Project Work at TU Wien is also gratefully acknowledged. 
The work of D.~P.\ is supported by the Austrian 
Science Fund FWF, project no.\ P26366. The work of F.~G.\ 
is supported by the Polish National Science Centre NCN through the OPUS project
nr.\ 2015/17/B/ST2/01625.

\end{document}